\documentclass{aa}
\usepackage{graphicx,amssymb,amsmath,natbib}
\bibpunct{(}{)}{;}{a}{}{,}
\begin{document}
   \title{Visual binaries among high-mass stars} \subtitle{An adaptive
optics survey of OB stars in the NGC\,6611 cluster }

\author{G. Duch\^ene\thanks{Visiting astronomer, Canada-France-Hawaii
Telescope Corporation operated by the National Research Council of
Canada, the Centre National de la Recherche Scientifique de France and
the University of Hawaii.}\inst{1,2} \and T. Simon\thanks{Visiting
astronomer, CFHT}\inst{3} \and J. Eisl\"offel\inst{4} \and
J. Bouvier\inst{1}}

\offprints{duchene@astro.ucla.edu}

\institute{Laboratoire d'Astrophysique de l'Observatoire de
  Grenoble BP53, F-38041 Grenoble cedex 9, France \and Division of
  Astronomy and Astrophysics, UCLA, Los Angeles, CA 90095-1562, USA
  \and Institute for Astronomy, University of  Hawaii, 2680 Woodlawn 
  Drive, Honolulu, HI 96822, USA
  \and Th\"uringer Landessternwarte Tautenburg, Sternwarte 5, D-07778
  Tautenburg, Germany }

   \date{Received 27 June 2001, accepted 19 September 2001}

   \abstract{We have searched for visual binaries with projected
   separations in the range 200--3000\,AU (0\farcs1--1\farcs5) among a
   sample of 96 stars in the massive young NGC\,6611 cluster, 60 of
   them being subsequently identified as high probability cluster
   members of mainly OB spectral type. This is the first visual binary
   survey among such a large and homogeneous sample of high-mass
   stars. We find an uncorrected binary frequency of 18$\pm$6\,\%
   over the surveyed separation range. Considering only binaries with
   mass ratios $q\geq0.1$, we find that OB stars in NGC\,6611
   host more companions than solar-type field stars. We derive mass
   ratios for the detected binaries from their near-infrared flux
   ratios and conclude that about half of the detected binaries have
   $q\lesssim0.2$, which does not contradict the assumption that
   companion masses are randomly drawn from the initial mass
   function. There is no evidence in our sample that wide-binary
   properties depend upon the mass of the primary star. The high
   frequency of massive binaries in a cluster as rich as NGC\,6611 and
   the lack of a strong mass dependence of their properties are
   difficult to reconcile with the scenario whereby massive stars form
   as the result of mergers of smaller stars. The canonical
   protostellar accretion scenario together with cloud fragmentation,
   on the other hand, can naturally explain most of the observed
   binary properties, although the very high stellar density in the
   protocluster is likely to require significant modification to
   that picture as well.
}

   \maketitle

\def\ng{NGC\,6611}

%

\section{Introduction}

Most detailed studies concerning star formation have focussed on
low-mass stars ($M\lesssim1\,M_\odot$), because they are more numerous
and relatively nearby
\citep{ghez93,leinert93,prosser94,simon95,petr98}. However, a growing
effort is underway to incorporate high-mass stars
($M\gtrsim10\,M_\odot$) in our understanding of this general
process. The potentially major impact of such massive and short-lived
stars includes their high photoionizing ultraviolet flux and
mechanical feedback on the interstellar medium through strong
mass-loss phenomena as well as dynamical interactions with
neighbouring low-mass objects.

The canonical inside-out collapse model for the formation of low-mass
objects \citep[e.g., ][]{shu87} cannot account for the formation of
massive stars due to the strong radiative flux from the central source
that eventually overcomes the spherical accretion flow; the critical
mass beyond which this occurs is $\sim10\,M_\odot$
\cite[]{beech_mitalas94}. The formation of such massive objects thus
requires another process in order to by-pass this hard limit. The
apparent tendency of OB stars to prefer the cores of dense stellar
clusters is not the result of rapid dynamic mass segregation
\citep{hillen_hartmann98,bonnell_davies98} since it is observed in
clusters which are younger than their relaxation times. Rather, a
dense environment appears to be a needed condition for the formation
of high-mass stars. A scenario that takes this factor into account has
recently been presented by \citet{bonnell98}, who suggested that these
objects form through tidally-induced mergers of intermediate mass
protostars in the early stages of the formation process. In this
framework, where dynamical interactions play a major role, it is
natural to study the properties of wide binaries in order to probe
this mechanism, as pairs with separations larger than a few tens of AU
are expected to be perturbed, if not disrupted, during close
encounters with other protostars. As opposed to this violent process,
the accretion paradigm could still be valid, provided it occurs in a
geometrically thin equatorial disk, which would not be blown away by
radiation pressure as easily as would a spherical envelope
\citep{nakano89}. Another alternative consists in spherical accretion
of highly dust-depleted material \citep{wolfire_cassinelli87}.

Because of their scarcity, OB stars are usually found far away from
the Sun (several hundreds of parsecs or more) so that direct imaging
surveys for visual binaries among the high-mass stars are strongly
constrained by resolution issues and have only rarely been attempted
so far. On the other hand, these stars offer many opportunities for
spectroscopic surveys because of their extreme
brightness. \citet{garmany80}, \citet{abt90}, and
\citet{morrell_levato91} have performed such surveys in the past and
have all concluded that the proportion of spectroscopic binaries among
massive stars is as large as that among solar-type stars, if not
significantly higher. With the advent of fast readout detectors,
speckle surveys of bright stars have also been conducted
\citep{mcalister93,mason98} and revealed a non-negligible fraction of
visual binaries among the OB stars, despite the limited dynamical
range of this technique. Overall, taking into account various surveys
probing different orbital period/separation ranges, it appears that OB
stars are often, if not always, found in multiple systems. Other
properties, such as orbital period and mass ratio distributions for
the OB binaries, are not as easy to estimate, due to the limited
sensitivity and the biases that are often inherent in inhomogeneous
samples.

One way to ensure both the homogeneity of the sample and a uniform
sensitivity to companions is to consider high-mass stars that are
located in a stellar cluster. Spectroscopic surveys for binarity have
already been conducted in several clusters \citep[for a review,
see][]{mermilliod_garcia01} or young stellar associations (Morrell \&
Levato 1991). Similar surveys for visual binaries among high-mass
stars have only been performed in the Orion Trapezium cluster by
\citet{preibisch99}, though in a sample comprised of just 13 OB stars. 
Other massive spectroscopic binaries in OB associations are known, but
were not discovered in the course of large-scale surveys
\citep[e.g.,][]{gieseking82}. To derive accurate statistical
properties for massive visual binaries, one must consider much richer
clusters that host several tens of OB stars.

\begin{figure}[t]
\includegraphics[width=\columnwidth,clip=true]{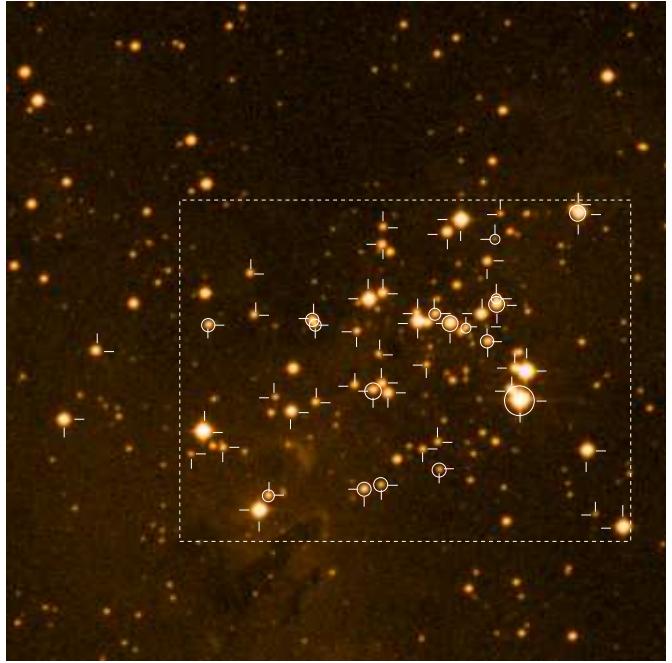}
\caption{\label{fig:zoom}Digital Sky Survey $10\arcmin\times10\arcmin$
image of the central part of the {\ng} cluster. Stars marked with
ticks are cluster members observed in this survey (including the faint
underexposed members), while circles identify all the newly discovered
pairs with separations smaller than 1\farcs5, both physical and likely
background companions. The dashed box, which encloses the core of the
cluster, contains about 75\,\% of the cluster members of spectral
type B0 or earlier. Another dozen members, including three binary
systems, were observed outside the core of the cluster; most of these
lie outside of this image.}
\end{figure}

{\ng} is the young, massive stellar cluster that is responsible for
the photoionization of the Eagle Nebula, famous for its Elephant
Trunks \citep{duncan20}. The first in-depth study of the cluster was
performed by \citet{walker61} in his pioneering work on star-forming
regions. He showed that {\ng} is an extremely young cluster containing
a pre-main sequence population of intermediate-mass objects, of
spectral type later than B5, together with several tens of higher mass
main sequence objects. Deeper studies of the cluster were subsequently
performed by \citet{the90}, \citet{hillenbrand93},
\citet{dewinter97} and \citet{belikov99,belikov00}. There is a general
consensus for a variable visual extinction, ranging from 1 to 5\,mag,
and an anomalous extinction law towards the cluster
\citep[e.g.,][]{dewinter97}, while other properties, such as the age
and distance of the cluster, remain somewhat uncertain. Estimates for
the distance to the cluster depend on the extinction law assumed and
have been given as 2.0\,kpc \citep{hillenbrand93}, 2.1\,kpc
\citep{mcbreen82,belikov99}, 2.3\,kpc \citep[][ average value from
anterior studies]{meaburn_white82} and 2.6\,kpc \citep{the90}. The age
of the cluster is also debated, but its extreme youth (at most a few
Myr) is evidenced by the presence of some very massive stars: the star
of earliest spectral type, W\,205, is an O5 main sequence object. The
median age of the cluster is of the order 2--3\,Myr
\citep{hillenbrand93,belikov00}, but the age spread within the cluster
is perhaps as large as 6\,Myr \citep{dewinter97,belikov00}. Throughout
this paper, we adopt the results of \cite{hillenbrand93}, i.e. an age
of 2\,Myr and a distance of 2\,kpc.

\begin{table*}
\centering
\begin{tabular}{lrcrll|lrcrll}
\hline
Walker & $K$ & memb. & $V$ & S.T. & ref & Walker & $K$ & memb. & $V$ &
S.T. & ref \\ 
number & & [\%] & & & & number & & [\%] & & & \\
\hline
25$^\S$ & 10.56 & 53 & 12.93 & B0.5 & 1 & 290$^\S$ & 10.82 & 52 &
12.14 & B2 & 2 \\ 
114$^\S$ & 13.70 & 32 & 16.11 & -- & -- & 296 & 10.16 & 95 &
11.78 & B2 & 3 \\
125$^\S$ & 8.53 & 79 & 10.01 & B1 & 1 & 299 & 12.75 & 55 & 14.46 &
B6 & 2 \\
150 & 8.28 & 86 & 9.85 & B0.5 & 1 & 300 & 10.94 & 16 & 12.69 &
B1.5 & 1 \\
162 & 12.99 & 28 & 15.27 & -- & -- & 305 & 9.77 &
84 & 13.51 & B1 & 4 \\
166 & 8.60 & 95 & 10.36 & O8.5 & 1 & 306 & 10.46 &
96 & 12.77 & B1.5 & 1 \\
175 & 7.07 & 95 & 10.09 & O5.5 & 1 & 307 & 11.08 &
76 & 14.18 & B1.5 & 1 \\
177 & 11.57 & 46 & 15.33 & -- & -- & 311 & 11.29 &
51 & 13.10 & B2.5 & 1 \\
188$^\S$ & 8.82 & 68 & 13.13 & B0 & 1 & 313 & 11.23 & 90 & 12.92 &
B4 & 2 \\
197 & 7.31 & 46 & 8.73 & O7 & 1 & 314 & 7.93 & 98 &
9.85 & B0 & 1 \\
205 & 6.80 & 41 & 8.18 & O5 & 1 & 322 & 11.41 & 26 &
13.68 & B8 & 1 \\
207 & 10.19 & 89 & 12.07 & B1 & 1 & 323 & 11.59 & 58
& 13.48 & B5 & 1 \\
210 & 9.95 & 92 & 11.68 & B1 & 1 & 336 & 11.51 & 88
& 13.29 & B3 & 2 \\
221 & 11.01 & 92 & 14.55 & B8 & 1 & 339 & 10.85 & 95 &
13.74 & B3 & 2 \\
223 & 9.48 & 89 & 11.20 & B1 & 1 & 343 & 8.91 & 83
& 11.72 & B1 & 1 \\
224 & 10.96 & 96 & 14.74 & -- & -- & 351 &
9.68 & 91 & 11.26 & B1 & 1 \\
227 & 10.55 & 89 & 12.85 & B1.5 & 1 & 360 &
10.76 & 74 & 14.48 & -- & -- \\
231 & 10.04 & 78 & 12.71 & B1 & 1 & 367 & 8.75 &
76 & 9.39 & O9.5 & 1 \\
235 & 7.93 & 96 & 10.98 & B1 & 3 & 371 & 11.19 & 72
& 13.44 & B0.5 & 1 \\
243 & 11.61 & 93 & 13.80 & B8 & 1 & 374 &
10.30 & 19 & 13.41 & F2 & 2 \\
246 & 6.66 & 88 & 9.46 & O7 & 1 & 388 & 11.81 & 90
& 13.70 & B6 & 2 \\
254 & 9.64 & 99 & 10.80 & B1 & 1 & 400 & 10.44 & 82 &
12.87 & B8 & 2 \\
259 & 9.20 & 87 & 11.61 & B0.5 & 1 & 401 & 7.59 & 46 &
8.90 & O8.5 & 1 \\
260 & 12.38 & 44 & 14.38 & -- & -- & 412$^\S$ & 7.26 & 34 & 8.18 &
O9.5 & 1 \\ 
262 & 11.67 & 68 & 13.96 & B7 & 2 & 468$^\S$ & 8.57 & 47 & 9.40
& B1 & 1 \\
267 & 11.47 & 87 & 13.13 & B2 & 2 & 469$^\S$ & 9.32 & 70 & 10.69 &
B0.5 & 1 \\
273 & 11.68 & 58 & 14.21 & A0 & 2 & 483$^\S$ & 9.51 & 72 & 10.99 &
B4.5 & 2 \\
276 & 11.39 & 69 & 13.74 & B5 & 2 & 489$^\S$ & 10.11 & 71 & 11.57 &
B7 & 2 \\
280 & 8.72 & 98 & 10.12 & O9.5 & 1 & 503$^\S$ & 7.86 & 40 & 9.75 &
B0--B0.5 & 3 \\
281 & 11.86 & 96 & 13.80 & A & 1 & 536$^\S$ & 10.21 & 39 &
11.46 & B1 & 1 \\
\hline
\end{tabular}
\caption{\label{tab:members}Cluster members observed (Main Sample)
together with our $K$-band photometry. A $^\S$ symbol denotes the 
stars that are located outside the central area of the cluster; membership
probabilities (col. 3 and 9) are from \citet{belikov99} and $V$
magnitudes from \citet{hillenbrand93}; 39 of these stars are ``very
likely'' ($3\,\sigma$) members following \citet{belikov99}. References
for the spectral classification: 1 -- \citet{hillenbrand93}; 2 --
\citet{dewinter97}; 3 -- \citet{the90}; 4 -- \citet{bosch99}.}
\end{table*}

\begin{table*}[t]
\centering
\begin{tabular}{lrcrlll|lrcrlll}
\hline
object & $K$ & memb. & $V$ & S.T. & ref. & note & object & $K$ &
memb. & $V$ & S.T. & ref. & note \\ 
 & & [\%] & & & & & & & [\%] & & & & \\
\hline
W\,136$^\S$ & 13.23 & 0 & 15.94 & -- & -- & n & W\,285 & 12.87 &
4 & 15.16 & -- & -- & n \\ 
W\,181 & 11.21 & 7 & 14.32 & B1.5 & 1 & n & W\,297
& 10.62 & 10 & 12.88 & B1.5 & 1 & n \\
W\,182 & 11.74 & 8 & 16.29 & -- & -- & n,f &
W\,301 & 10.42 & 92 & 15.49 & B2 & 1 & f \\
W\,198 & 11.56 & 0 & 13.21 & -- & -- & n & W\,309
& 11.16 & 0 & 15.49 & -- & -- & n \\ 
W\,202 & 11.19 & 0 & 14.40 & B3 & 2 & n & W\,321 &
9.70 & 0 & 15.33 & -- & -- & n \\ 
W\,209 & 12.79 & 0 & 15.81 & -- & -- & n,f &
W\,335 & 12.71 & 0 & 16.58 & -- & -- & n,f \\ 
W\,213 & 9.48 & 0 & 14.18 & A7--F9 & 2 & n &
W\,353 & 14.45 & 0 & 16.55 & -- & -- & n,f \\
W\,226 & 13.22 & 14 & 15.88 & -- & -- & f & W\,355
& 13.49 & 0 & 15.71 & -- & -- & n,f \\ 
W\,229 & 13.38 & 30 & 16.20 & -- & -- & f &
W\,362 & 13.41 & 2 & 15.89 & -- & -- & n \\ 
W\,237 & 12.41 & 2 & 15.17 & -- & -- & n & W\,364
& 11.36 & 74 & 13.44 & -- & -- & f \\ 
W\,240 & 12.54 & 6 & 14.56 & B8 & 2 & n & W\,366 &
11.00 & 0 & 14.27 & -- & -- & n \\
W\,245 & 9.79 & 1 & 13.61 & B6 & 2 & n & W\,396 &
10.39 & 3 & 14.03 & F9--G2 & 2 & n,f \\
W\,252 & 11.44 & 0 & 13.90 & -- & -- & n & W\,455$^\S$ & 10.53 &
7 & 12.11 & B7 & 2 & n \\
W\,257 & 12.51 & 0 & 15.59 & -- & -- & n & W\,460$^\S$ & 12.48 &
0 & 15.36 & -- & -- & n,f \\ 
W\,258 & 12.65 & 0 & 15.43 & -- & -- & n & W\,487$^\S$ & 12.40 &
12 & 15.00 & -- & -- & n \\ 
W\,266 & 9.61 & 0 & 14.35 & F8 & 2 & n & KS\,29617
& 11.47 & 0 & 16.30 & -- & -- & n,f \\
W\,270 & 12.83 & 0 & 15.55 & -- & -- & n,f &
KS\,29710 & 11.60 & 37 & 11.48 & -- & -- & f \\
W\,275 & 10.64 & 0 & 12.12 & B1.5 & 1 & n & KS\,29797$^\S$ &
14.46 & 4 & 16.03 & -- & -- & n \\
\hline
\end{tabular}
\caption{\label{tab:nonmembers}List of additional stars observed in
our programme. The notes in col. 7 and 14 indicate the reason why the
star is excluded from our sample: ``n'' means that the star is not a
likely cluster member, while ``f'' indicates that the exposure time used
during the observation was insufficient to ensure a good detection
limit. KS numbers are from \citet{kharchenko_schilbach95}; other
columns have the same meaning as in Table\,\ref{tab:members}.}
\end{table*}

The first systematic, though limited, search for binaries in {\ng} was
performed by \citet{bosch99}, who conducted a spectroscopic survey of
the 10 earliest type stars in the cluster. In this paper, we present
the first high-angular resolution imaging binary survey of the OB
population of {\ng}, which is an extension of our low-mass binary
surveys in young clusters \citep{bouvier97,duchene99ic,bouvier01}. We
aim at (i) estimating some of the binary properties with which we can
then confront the collisional formation model of high-mass stars, and
(ii) comparing directly these properties to those that have been
previously estimated for lower mass binaries. The sample and
observations are described in Sect.\,\ref{sec:obs};
Sect.\,\ref{sec:res} describes our results and compares them to
previous OB multiplicity surveys. Implications for the scenarios of
massive star formation are then considered in
Sect.\,\ref{sec:discus}. Finally, Sect.\,\ref{sec:concl} summarizes
the main results of our study.


\section{Sample selection and observations}
\label{sec:obs}


\subsection{Sample definition}

Assessing membership for stars in {\ng} is an especially crucial issue
in view of the location of the cluster near the Galactic plane
($l\sim17\deg$, $b\sim0\fdg8$), its large distance, and its moderate
intracluster extinction. Numerous foreground low-mass dwarfs as well
as bright background giants are likely to be projected on the cluster
and thus confused with true cluster members. Furthermore, the
intrinsic proper motion of the cluster is very small
\citep[$\sim~2.5$\,mas/yr, ][]{belikov99} and not well
differentiated from that of the field stars. Combining proper motion
measurements with the spatial location of the stars, \citet{belikov99}
estimated individual membership probabilities for all stars brighter
than $V=16.8$. Most of the stars with spectral type earlier than B5
were thus confirmed as cluster members, providing a robust sample for
a statistical study of binarity in this cluster. \citet{belikov00}
further used optical photometry data to reassess cluster membership.
Statistically, about 75--80\,\% of their sample was confirmed as
belonging to the cluster. On the other hand, the low-mass T\,Tauri
population of {\ng} is totally unknown due to the distance to the
cluster and the extreme crowding of the field for fainter stars.

Our sample, drawn from the lists of \cite{walker61} and of
\cite{kharchenko_schilbach95}, contains 59 OB stars as well
as about twenty intermediate-type objects. Most of these objects (66
out of 82 targets) are located in the central area of the cluster,
which contains most of the O-type stars of the cluster (this area is
defined in Figure\,\ref{fig:zoom}). An additional fourteen objects
from the lists mentioned above were detected in our images, but were
not exposed long enough to ensure a satisfactory companion
detectability, and so are not considered further.

We use the membership probabilities derived by \citet{belikov99},
which were published after our observations were completed, to
identify cluster members. These authors set two probability thresholds
at 14\,\% and 61\,\%, which correspond respectively to ``likely''
($2\,\sigma$) and ``very likely'' ($3\,\sigma$) cluster members. In
the following, our Main Sample consists of all sources having
membership probabilities larger than 14\,\% and that were bright
enough to ensure the detection of a 6\,mag flux ratio companion at
1{\arcsec}; it contains 51 OB stars and a few later-type objects. The
60 members that form our Main Sample are listed in
Table\,\ref{tab:members}, while the remaining 36 targets are presented
in Table\,\ref{tab:nonmembers}, including ``faint'' stars. The
observed members located in the core of the cluster are indicated in
Figure\,\ref{fig:zoom}, as are the newly discovered binary
systems. Only two members located in the core and brighter than
$V=14.5$ were not observed in this survey (W\,349 and W\,402); both
are late G-type (super)giants according to \citet{dewinter97}.


\subsection{Observations and data reduction}

We used the adaptive optics system, PUEO \citep{rigaut98}, at the
Canada-France-Hawaii Telescope with the 1024$\times$1024 near-infrared
detector KIR \citep{doyon98} during four nights in June and July
1998. The pixel size is about 0\farcs035, yielding a total field of
view of 36\arcsec. Each target employed for wavefront sensing was
observed in a four-step dithered pattern, which increased the field of
view to a final image size of about 50\arcsec\ and generally resulted
in the detection of several cluster members per field. Total
integration varied from 3 seconds to 15 minutes depending on the
source brightness and field crowding. Data reduction included all the
usual steps for near-infrared images (sky subtraction, flat-fielding
and shift-and-add), and was performed with standard IRAF\footnote{IRAF
is distributed by the National Optical Astronomy Observatories, which
is operated by the Association of Universities for Research in
Astronomy, Inc., under contract to the National Science Foundation}
routines.

\begin{figure*}[t]
\includegraphics[width=\textwidth]{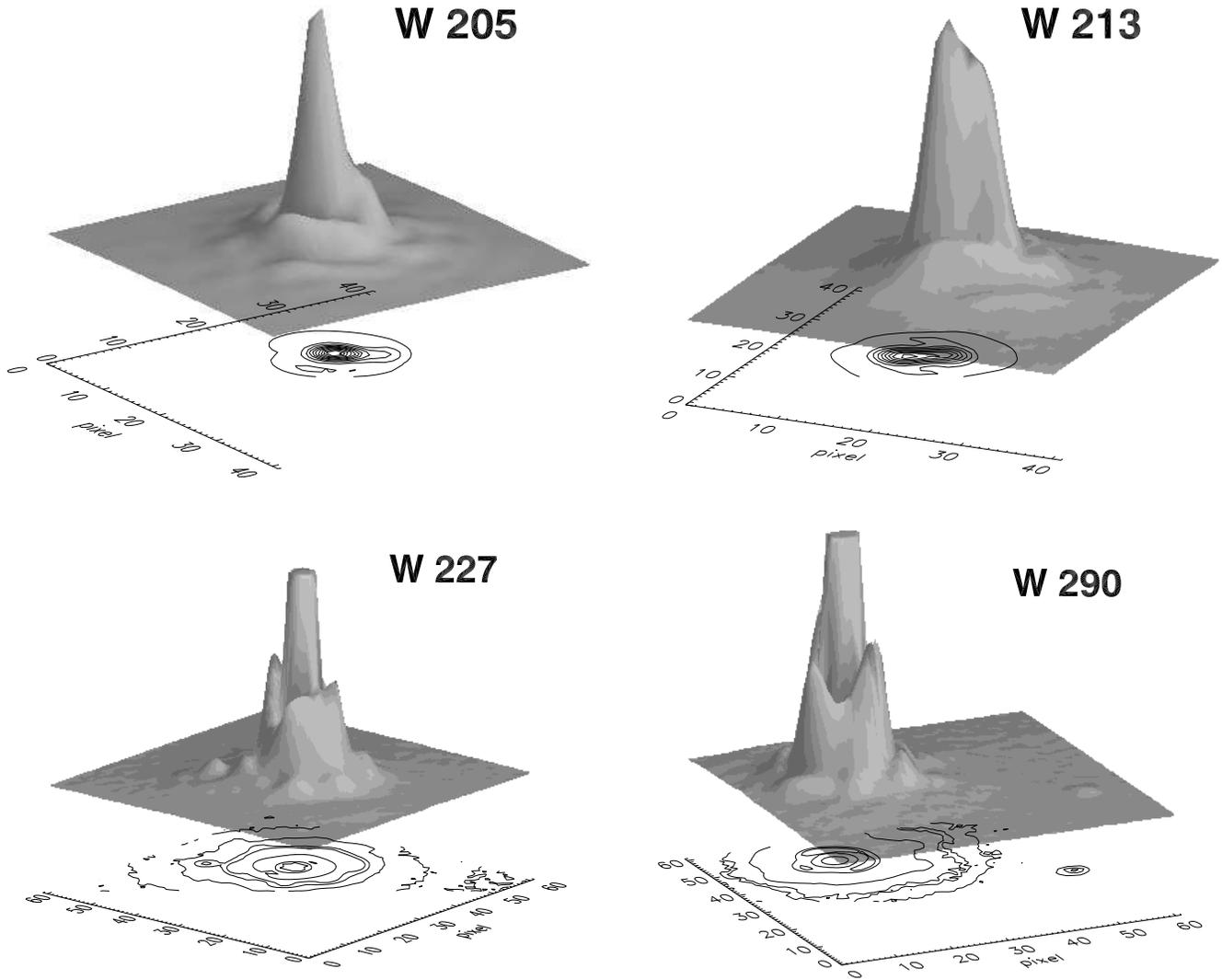}
\caption{Surface and contour plots of some of the binaries detected in
this survey,illustrating the resolving power of our observing
technique. The upper two images are 1\farcs4 on a side and represents
the tightest pairs that we found, while the lower ones are 2\farcs1
wide and illustrate large flux ratios in wider systems. All images
were obtained in the $K$-band. Note the excellent adaptive optics
correction in the image of W\,290, in which the first Airy ring is
almost azimuthaly symetric as well as the clear detection of the
companion although the binary flux ratio is $\Delta K\approx
5.9$\,mag. On the other hand, the companion to W\,227 (to the left of
the primary in this image) appears barely above the noise in the
imperfect adaptive optics halo.}
\label{fig:exemple}
\end{figure*}

To deal with demanding contrast issues, we conducted the survey in the
$K$-band, since high-mass stars fade much more rapidly at longer
wavelengths than do their putative low mass secondaries. Our images
revealed a large number of stars, up to several hundred in each field,
due to the low galactic latitude of the cluster. The tight pairs
detected in the raw images at the telescope were immediately observed
in the $J$ and $H$ bands. Some of the binaries identified in this
survey are shown in
Figure\,\ref{fig:exemple}. Figure\,\ref{fig:detect} presents a plot of
the $K$-band differential photometry for the companions that were
detected at $K$, as well as our detection limit for that
wavelength. The latter is an azimuthally averaged value, which has
been estimated by adding artificial faint companions to the images of
single stars. Figure\,\ref{fig:detect} also shows the impact of a
change in image quality on the detectability of companion stars. Such
a change mostly affects the inner 0\farcs5, where the
diffraction-limited part of the stellar point spread function (PSF)
dominates the flux. Our detection limit has further been empirically
confirmed by combining all companions detected with the same
instrumentation in the course of our survey of young clusters
\citep{bouvier_duchene01}.

Some of the companions detected in the $K$-band were undetected in the
$J$-band, and sometimes even in the $H$-band (e.g., W\,275 and
W\,297). This is the result of two complementary effects: (i) the
intrinsic brightness difference between the two stars increases
strongly towards shorter wavelengths, and (ii) the adaptive optics
correction degrades when observing at the shorter wavelengths,
especially in the $J$-band, thereby increasing the flux in the halo of
the primary. We were able to estimate minimum flux ratios in those
cases by subtracting the PSF of the primary and then extracting a flux
limit by measuring the noise level at the known location of the
secondary.

Aperture photometry was performed through 2{\arcsec}-radius apertures
and then absolute calibration in the CIT system was obtained by
comparison to the \citet{hillenbrand93} photometry of the cluster
area. Based on the comparison of 40 to 80 independent measurements in
each filter, the standard deviation between the two datasets is
$\lesssim0.1$\,mag, including possible long-term variability for some
stars. Binary flux ratios, separations and position angles were
usually obtained through PSF fitting and/or aperture photometry for
wide enough systems. Flux ratio uncertainties are estimated by
comparing various images of the same system; it appears to be
0.03\,mag in both cases. We employed a Richardson-Lucy deconvolution
algorithm for the tighest pairs and subtracted the PSF of the primary
star for those companions that were just barely above our detection
limit. The two methods in principle require perfect knowledge of the
PSF; both yield uncertainties of the order of 0.1\,mag. Typical
uncertainties in our astrometric measurements are about 0\farcs005 for
the separation and 0\fdg2 for the position angle on the sky, half
coming from the absolute calibration, which was derived from
observations of several well-known binaries from the Index Catalogue
of Double Stars \citep{vandessel_sinachopoulos93}. In the few cases
where the secondary was barely above our detection limit, these
uncertainties can reach 0\farcs01 and up to 4{\degr} for the tightest
pairs. Five binaries are affected by these larger uncertainties,
namely, W\,205, W\,213, W\,223, W\,224 and W\,260.


\section{Results}
\label{sec:res}

Because most of our images contain large numbers of sources, we cannot
{\it a priori} consider a random pair of stars as being a physical
binary system; we first have to set a criterion to exclude very wide
pairs and then check that the remaining systems are indeed physically
linked. We first adopt a 1\farcs5 upper limit for considering pairs,
both because this corresponds to a fairly large physical separation
(3000\,AU) at the distance of {\ng} and because the average distance
to the nearest neighbour down to $K\approx18.5$ is about 1\farcs4 in
our most crowded field. Photometric and astrometric measurements for
all detected pairs are summarized in Table\,\ref{tab:bin} where seven
candidate background companions, which are identified in this section,
are marked.

Before we discuss statistical properties of the high-mass binary
population of {\ng}, it is important to determine which of these pairs
indeed form bound systems (Sect.\,\ref{subsec:bound}). In this
section, we also estimate the binary mass ratios
(Sect.\,\ref{subsec:q}) and derive a lower limit for the total binary
frequency of OB stars in {\ng} and compare this value to previous
binary surveys (Sect.\,\ref{subsec:totbf}). Finally, we summarize
other important trends (Sect.\,\ref{subsec:binprop}).


\subsection{Bound companions}
\label{subsec:bound}

We must consider two aspects to confirm the physical status of the
identified pairs. First, the two stars must be cluster members located
at the same depth in the cloud, as opposed to two unrelated members
that merely appear close together on the sky due to projection
effects. When this is secured (Sect.\,\ref{subsubsec:nir}), we have to
check that such wide pairs form bound systems. The latter point can be
argued from studies of lower-mass stars: 3000\,AU is a safe upper
limit for separations of solar-type binaries
\cite[e.g.,][]{duq_may91}. Furthermore, \citet{abt88} estimated a
maximum separation threshold of about 20000\,AU for B5 primaries and
it is likely that more massive primaries sustain even wider systems.
We thus infer that any cluster star physically located within 3000\,AU
of another high-mass cluster star can be safely interpreted as a bound
companion. However, two cluster members can be projected next to each
other on the sky while being at different depths in the cluster. A
quantitative estimate of such cases is presented in
Sect.\,\ref{subsubsec:prob}, simultaneously considering non-members
as possible false companions.

\begin{figure}[t]
\includegraphics[angle=270,width=\columnwidth]{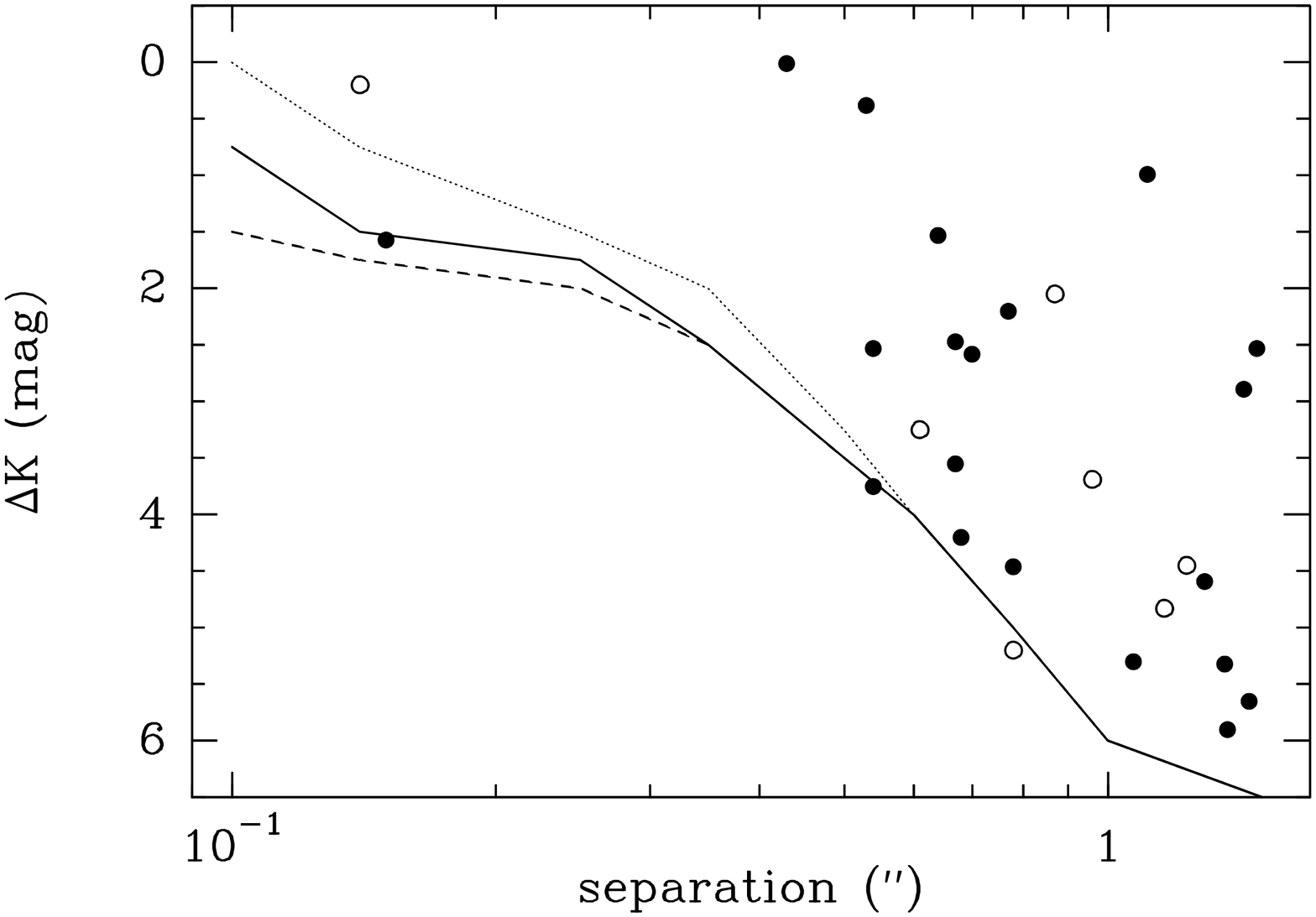}
\caption{Detected companions around our targets and overall detection
limit of our survey. Plotted is the $K$-band flux ratio as a function
of the binary separation. Filled circles denote close companions to
cluster members (including around the ``faint'' members W\,226 and
W\,364) found in this survey, while open circles denote
non-members. The solid line is our typical detection limit while
dashed and dotted curves correspond respectively to the best and worst
image quality achieved in our survey.}
\label{fig:detect}
\end{figure}

\subsubsection{Near-infrared properties of detected companions}
\label{subsubsec:nir}

\begin{figure}[t]
\includegraphics[angle=270,width=\columnwidth,clip=true]{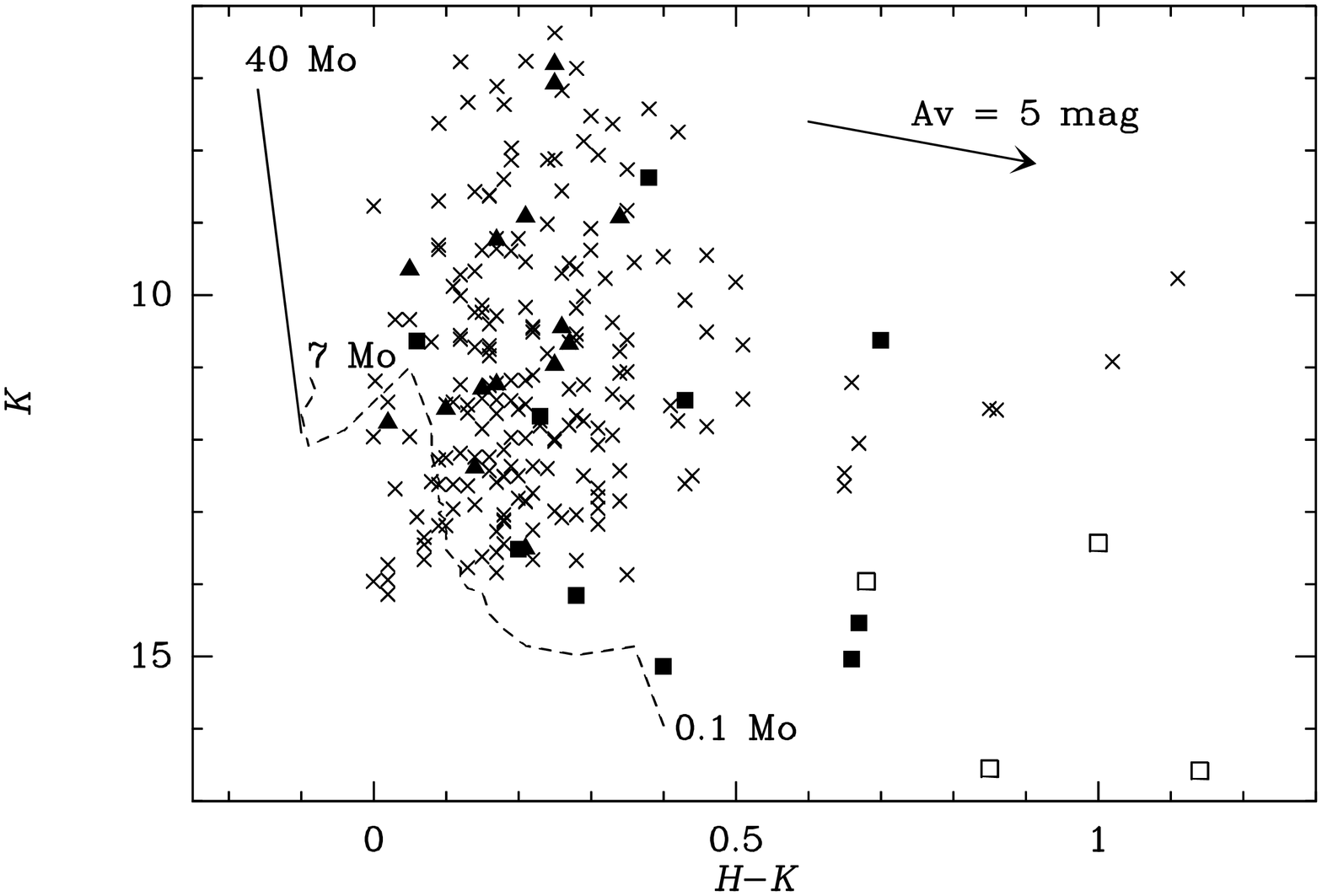}
\caption{Near-infrared colour-magnitude diagramme for {\ng}. All likely
(2\,$\sigma$) members from \citet{belikov99} are presented as crosses,
using photometry from \citet{hillenbrand93}. Filled triangles and
squares represent member primaries and secondaries of binaries that
were detected in our Main Sample. Suspected background secondaries are
shown as open squares; the background companion to W\,343, with its
$H-K\approx1.8$\,mag colour index, is not plotted here. The solid line
is a linear fit to the synthetic colours of \citet{schaller92} for
high mass stars ($M>4M_\odot$), while the dashed curve shows the
pre-main sequence evolutionary models of \citet{siess00} for stars
with masses $0.1\,M_\odot<M<7\,M_\odot$; both models correspond to a
2\,Myr cluster age. The local maximum at $H-K\approx0.05$
corresponds to a mass of $3\,M_\odot$, above which stars have already
reached the ZAMS. Deuterium burning in very low-mass stars is
responsible for the plateau at $0.2<H-K<0.4$. A typical extinction
vector following the canonical law of \cite{rieke_lebof85} is also
shown.}
\label{fig:cmd}
\end{figure}

First, we infer membership for the putative secondaries by plotting
all components of the detected binaries in a near-infrared $(K,H-K)$
colour-magnitude diagramme. As can be seen in Figure\,\ref{fig:cmd},
the primaries of the known members of the cluster form a vast cloud of
points. About half the secondaries also lie in the same locus,
suggesting that they really are cloud members. However, a few
companions are isolated from other members in this diagramme. They
could be either highly extincted low-mass members of the cluster or
else heavily reddened objects in the background that are seen through
the molecular cloud. One way to distinguish between these two
possibilities is to compare the apparent extinction needed to bring
the primaries and their secondaries back onto the same theoretical
isochrone, assuming both stars have the same age. Since both stars are
equally embedded in the molecular cloud, one expects both components
of a system to show a similar extinction, unless one of them suffers a
significant circumstellar extinction. Consequently, secondaries which
appear much more extincted than their primaries are serious background
giant candidates. Five of the seven such candidates identified in
Table\,\ref{tab:bin} are excluded by virtue of this criterion; in each
of these cases, the extinction $A_V$ towards the secondary appears to
be at least 5\,mag larger than that along the line of sight to the
primary.

In addition, when a ``companion'' is detected in the $J$-band, one
can also use the $J-H$ colour index to confirm its status:
\citet{hillenbrand93} suggested that objects with $J-H>1.6$ likely lie
beyond the cluster. Among the five sources mentioned above, two
display $J-H>1.6$ (W\,243 and W\,339), while the other three are not
detected in the $J$-band, but have suspicious $H-K$ colour indexes
($H-K>0.85$). Since none of the other companions shows such extremely
red colours, we conclude that these five objects are all very likely
to be background giants\footnote{A. Ghez and C. McCabe subsequently
obtained a low-resolution long slit $K$-band spectrum of W\,243 with
NIRPSEC behind the adaptive optics system at W. M. Keck Observatory,
with the slit oriented along the binary. The spectrum of the secondary
revealed no significant feature, suggesting a spectral type F or
earlier, inconsistent with its photometric mass estimates (see
Table\,\ref{tab:secmass}). This confirms the classification of this
companion as a background star.}. Beyond our Main Sample, the
companion to W\,226 is also a likely background object while W\,364
seems to be a physical system. We also emphasize that both of these
criteria help identifying background stars but do not provide any
evidence for/against the presence of projected foreground objects.

\subsubsection{Statistical contamination by field stars}
\label{subsubsec:prob}

We consider here the possibility that some of the close pairs simply
result from random pairing of field stars. To do so, we first compute
the average spatial density of objects that are at least as bright as
the companion in the image. We then estimate the probability of
finding one such object at a distance from the primary at most equal
to that of the companion.

\begin{table*}[t]
\centering
\begin{tabular}{lrrrrrlllcl}
\hline
Walker & sep. & P.A. & $K_A$ & $H_A$ & $J_A$ & $\Delta K$ & $\Delta H$
& $\Delta J$ & $N(K\leq K_B)$ & $P_{bound}$ \\
number & [\arcsec] & [\degr] & & & & & & & in field & \\
\hline
\hline
\multicolumn{10}{c}{Main Sample binaries}\\
\hline
25 & 1.432 & 288.2 & 10.56 & -- & -- & 2.89 & -- & -- & 7 & 0.985 \\
175 & 0.672 & 231.1 & 7.07 & 7.32 & -- & 3.55 & 4.0$\pm$0.1 & -- &
1 & 0.999 \\
188 & 0.543 & 227.7 & 8.92 & 9.26 & 9.66 & 2.53 & 2.62 & 2.80 &
1 & 0.999 \\
205 & 0.148 & 320.0 & 6.80 & 7.05 & -- & 1.57$\pm$0.1 & 1.7$\pm$0.1 &
-- & 1 & 0.999 \\
223$^\dagger$ & 1.081 & 295.0 & 9.49 & 10.07 & 9.95 & 5.3$\pm$0.1 &
5.8$\pm$0.2 & $>6.1$ & 48 & 0.955 \\ 
224$^\dagger$ & 0.779 & 321.9 & 10.99 & 11.73 & 11.97 & 4.46 & 4.59 &
$>4.5$ & 79 & 0.963 \\ 
227 & 0.538 & 29.8 & 10.58 & -- & -- & 3.75 & -- & -- & 46 & 0.991 \\
243$^\P$ & 0.770 & 224.5 & 11.76 & 11.78 & 12.12 & 2.20 & 2.86 &
4.5$\pm$0.1 & 28 & 0.975 \\
254 & 1.107 & 184.9 & 9.64 & 9.69 & 9.82 & 0.99 & 1.00 & 1.05 &
3 & 0.997 \\
260$^\P$ & 0.676 & 152.6 & 12.38 & 12.52 & 12.78 & 4.2$\pm$0.1 &
5.2$\pm$0.1 & $>6.0$ & 107 & 0.962 \\
267 & 0.701 & 136.3 & 11.57 & 11.67 & 11.80 & 2.58 & 2.76 &
3.9$\pm$0.1 & 48 & 0.974 \\
290$^\S$ & 1.373 & 235.3 & 10.82 & -- & -- &  5.9$\pm$0.1 & -- & -- &
134 & 0.793 \\
299 & 0.433 & 94.7 & 13.50 & 13.71 & 14.09 & 0.01 & 0.00 & -0.03 &
13 & 0.998 \\
311 & 0.530 & 254.3 & 11.29 & 11.44 & 11.80 & 0.38 & 0.46 & 0.47 &
5 & 0.998 \\
313$^\P$ & 1.364 & 297.0 & 11.23 & 11.40 & 11.61 & 5.32 &
6.0$\pm$0.2 & $>6.5$ & 123 & 0.783 \\
339$^\P$ & 0.670 & 306.8 & 10.96 & 11.21 & 11.74 & 2.47 & 3.22 &
$>4.5$ & 15 & 0.992 \\
343$^\P$ & 1.449 & 89.6 & 8.91 & 9.12 & 9.52 & 5.65 & 7.3$\pm$0.3 &
$>6.5$ & 41 & 0.921 \\
400$^\S$ & 1.288 & 320.3 & 10.44 & 10.70 & 11.04 & 4.59 & 4.99 &
5.9$\pm$0.2 & 51 & 0.900 \\
\hline
\hline
\multicolumn{11}{c}{Binaries among ``faint members''}\\
\hline
226$^\P$ & 0.643 & 346.1 & 13.46 & 13.40 & 13.87 & 1.53 & 2.18 &
3.9$\pm$0.2 & 52 & 0.986 \\
364 & 1.484 & 267.8 & 11.36 & 11.58 & 11.75 & 2.53 & 2.87 & 3.62 &
16 & 0.960 \\
\hline
\multicolumn{11}{c}{Non-member primaries}\\
\hline
213 & 0.139 & 298.7 & 10.14 & 10.49 & 11.19 & 0.2$\pm$0.1 &
0.2$\pm$0.1 & 0.2$\pm$0.1 & 1 & 0.999 \\
237 & 1.156 & 273.3 & 12.41 & -- & -- & 4.83 & -- & -- & 207 &
0.778 \\
270 & 0.872 & 6.6 & 13.36 & 13.39 & 13.74 & 2.05 & 2.7$\pm$0.2 &
$>3.2$ & 94 & 0.927 \\
275 & 1.233 & 253.2 & 10.64 & 10.75 & 10.92 & 4.45 & $>4.6$ & $>4.6$ &
94 & 0.879 \\
297 & 0.775 & 120.4 & 10.62 & 10.69 & 11.01 & 5.2$\pm$0.1 &
$>6.0$ & $>6.0$ & 102 & 0.973 \\
362 & 0.959 & 261.3 & 13.41 & 13.60 & 14.03 & 3.69 & 4.6$\pm$0.2 &
$>4.4$ & 148 & 0.862 \\
487 & 0.611 & 323.8 & 12.45 & -- & -- & 3.25 & -- & -- & 34 & 0.987 \\
\hline
\end{tabular}
\caption{\label{tab:bin}Observed properties of all detected binaries
in {\ng} with separations smaller than 1\farcs5. The last two columns
give the number of stars brighter than the secondary in the same field
and our empirical estimate of the probability that a system is
physically bound. A $^\P$ symbol indicates that the apparent companion
is a likely background giant given its apparent extinction (see
section\,\ref{subsec:bound}) while a $^\S$ flags those companions with
suspiciously low mass (section\,\ref{subsec:q}); all other systems
have probabilities of being real which are above 95.5\,\% (2\,$\sigma$
confidence level). The field containing the two systems indicated by a
$^\dagger$ symbol were observed under non-photometric conditions and
we used the \cite{hillenbrand93} unresolved photometry together with
our observed flux ratios for the analysis.}
\end{table*}

Equivalently, we estimate here $P_{bound}$, the complementary quantity
to this probability, that no random star brighter than the secondary
($K_p\leq K\leq K_s$) is detected within the binary separation $r_b$
from the primary, as being $P(N=0,\mu=\overline{N})$ derived from
Poisson statistics. Our detection limit for companions is explicitely
taken into account as we use the following expression for
$\overline{N}$: $$\overline{N} = \sum_{K = K_p}^{K_s}{n_K W_K} $$
\noindent
where $n_K$ and $W_K$ are respectively the cluster surface density in
the considered field and the detectability area of stars at a given 
magnitude $K$, computed as follows: 
$$ W_K = \int_{0''\!\!\!.\,1}^{r_b}{V_K(r)\times 2\pi r\,dr}. $$
\noindent
We are assuming that the visibility $V_K(r)$ is simply 
$$ V_K(r) = H(r - r_K) = \left \{ \begin{array}{ll} 0 & \mbox{if $r <
r_K$}\\ 1 & \mbox{if $r \geq r_K$} \end{array} \right. $$
\noindent
where $r_K$ corresponds to the minimum binary separation at which,
according to Figure\,\ref{fig:detect}, a secondary of infrared
magnitude $K$ is detectable in our images. Therefore, it follows that
$ W_K = \pi\, (r_b^2 - r_K^2).$ The derived values for $P_{bound}$ are
listed in the last column of Table\,\ref{tab:bin}.

Although most systems appear likely bound ($P_{bound}>95.5\,\%$,
corresponding to a 2\,$\sigma$ confidence level, for 15 out of 18
binary members), the probability that all pairs are physically linked
is only 41\,\%, thus urging for a deeper analysis. For six pairs, the
probabilities of being physical are higher than 99.7\,\% ($3\,\sigma$
confidence level).  These are very likely to be real systems. It is
also likely that some systems with lower probabilities are real as
well, even though one of the physically unrelated pairs identified
above (W\,339) has an uncomfortably high probability (99.2\,\%) of
being real from a purely statistical analysis. This is a reminder that
as much information as possible should be used to ascertain membership
and that one should not place reliance on statistical arguments alone
in individual cases. If we consider all binaries down to a $2\,\sigma$
threshold while excluding those rejected as projected systems
(including from Sect.\,\ref{subsec:q}), we are left with 11 candidate
cluster binaries in our Main Sample. Both this number and the six
``very high probability'' binaries are considered in further sections.


\subsection{Masses of the detected companions}
\label{subsec:q}

In this section, we use current theoretical evolutionary models to
estimate the mass of the detected secondaries which, in turn, allows
us to derive binary mass ratios. The resulting secondary masses, 
derived from the 2\,Myr mass-luminosity relation constructed below, 
are summarized in Table\,\ref{tab:secmass}. In principle, the mass 
of every cluster member, including the primaries, can be derived 
from this mass-luminosity relation and our near-infrared photometry.  
However, the variable and anomalous extinction towards the cluster 
prevents this, and so we are obliged to take a more indirect route. 

With regard to the primaries, those of known spectral type are easily
handled by using the theoretical relation between mass and effective
temperature/spectral type for massive stars. We have used models from
\citet{schaller92} to estimate masses for stars already on the main 
sequence, together with spectral type-effective temperature conversion
tables from \citet{bohm81}. Pre-main sequence evolutionary models are
needed for stars with masses below 4\,$M_\odot$, i.e., those of
spectral type later than B5, since they are still contracting. For
those objects, we have used models from \citet{siess00}, which extend
from 7\,$M_\odot$ down to 0.1\,$M_\odot$. Both sets of models are in
good agreement over the common 4--7\,$M_\odot$ mass range. In the last
stages before reaching the main sequence, stars undergo a rapid phase
of contraction and exhibit a strong dependence of effective
temperature on mass: a 3\,$M_\odot$ star has a spectral type G0, while
a 3.5\,$M_\odot$ star is already at B7, according to the
\citet{siess00} models.  Consequently, we did not estimate the exact
mass of the primaries with spectral types later than B5; rather a
3--4\,$M_\odot$ range was considered. The primaries of W\,224 and
W\,260 have unknown spectral types, so that the mass ratio of these
systems cannot be estimated.

\begin{figure}[t]
\includegraphics[angle=270,width=\columnwidth,clip=true]{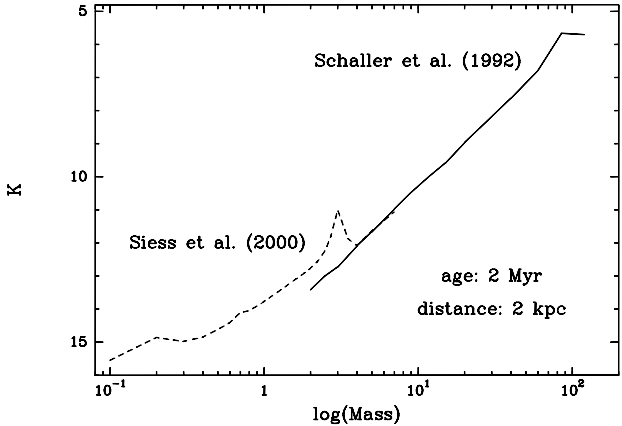}
\caption{Theoretical $K$-band unreddened magnitude-mass relationship for
2\,Myr-old stars, as used in this work. For masses below 7\,$M_\odot$,
we used the Siess et al. (2000) evolutionary models which take
pre-main sequence evolution into account. For higher masses, the
models of Schaller et al. (1992) were used in conjunction with
bolometric corrections from Kenyon \& Hartmann (1995) and
B\"ohm-Vitense (1981).}
\label{fig:masslum}
\end{figure}

To derive masses for the companions, we first constructed theoretical
$J$-, $H$- and $K$-band mass-luminosity relations from the
\citet{siess00} and \citet{schaller92} models. The $K$-band relation
is presented in Figure\,\ref{fig:masslum}. Given the mass of the
primary implied by its spectral type (as described in the previous
paragraph), we estimated the ``theoretical'' brightness of the primary
through the above relation, we then used the observed flux ratio to
infer the ``theoretical'' brightness of the companion, and then we
re-entered the above relation to obtain the mass of the secondary
star. The underlying assumption to this method, of course, is that
both stars are identically embedded, i.e., the extinction is mostly
due to the molecular cloud itself.

For real companions, the masses estimated independently from different
filters should (and apparently do) agree well with each other, while
significant variations and/or very low masses are found in the case of
likely background companions, because of their heavier extinction and
extremely red colours. The slight changes in mass with wavelength
listed for W\,223 and W\,267 in Table\,\ref{tab:secmass} can be
explained by photometric uncertainties and/or moderate differential
extinction while that observed for W\,339 is significant. In two
systems not already rejected, W\,290 and W\,400, the companions are so
faint that they are brown dwarf candidates {\it if} they belong to the
cluster. Alternatively, both companions may be background giant stars,
which is also suggested by their low probabilities of being real
companions, below the $2\,\sigma$ confidence level. Both systems have
been flagged as candidate background companions in
Table\,\ref{tab:bin}. The companion to W\,400 appears only slightly
more extincted than its primary by $\Delta A_V\sim$2--3\,mag while
W\,290 was only detected in the $K$-band, therefore its colours could
not be used to reject it as a non-physical system.

\begin{table}[t]
\centering
\begin{tabular}{llllll}
\hline
star & $M^A$ & $M^B_K$ & $M^B_H$ & $M^B_J$ & $q=\frac{M_B}{M_A}$ \\
 & [$M_\odot$] & [$M_\odot$] & [$M_\odot$] & [$M_\odot$] & \\
\hline
\hline
W\,25 & 12.5 & 2 & -- & -- & 0.16 \\
W\,175 & 33.5 & 2--5.5 & 2.7--4.6 & -- & 0.08--0.14 \\
W\,188 & 12.5 & 2.4 & 2.4 & 2.6 & 0.2 \\
W\,205 & 38 & 16.5 & 16.0 & -- & 0.42 \\
W\,223 & 10 & 0.10 & 0.13 & -- & $\sim$0.01 \\
W\,227 & 8.8 & 0.7 & -- & -- & 0.08 \\
W\,254 & 10 & 6 & 6 & 6 & 0.6 \\
W\,267 & 7.7 & 1.3 & 1.3 & 1.0 & 0.15 \\
W\,299 & 3--4 & {\it 3--4} & {\it 3--4} & {\it 3--4} & {\it 1.0} \\
W\,311 & 6.4 & 2--5 & 2.8--5 & 3--5 & 0.47--0.78 \\
\hline
\hline
\multicolumn{6}{c}{Background companions from Sect.\,\ref{subsec:bound}}\\
\hline
W\,243 & 3--4 & 1.3--0.7 & 0.9--0.45 & 0.2--0.15 & -- \\
W\,313 & 4 & $<$0.1 & $<$0.1 & $<$0.1 & -- \\
W\,339 & 5.8 & 0.95 & 0.66 & -- & -- \\
W\,343 & 10 & $<$0.1 & $<$0.1 & $<$0.1 & -- \\
\hline
\hline
\multicolumn{6}{c}{Background companions from Sect.\,\ref{subsec:q}}\\
\hline
W\,290 & 7.7 & $<$0.1  & -- & -- & --  \\
W\,400 & 3--4 & $<$0.1 & $<$0.1 & $<$0.1 & -- \\
\hline
\end{tabular}
\caption{Masses of both components in all binaries of our Main Sample
with a primary of known spectral type. The three estimates for the
secondaries were
independently obtained in each filter. For some primaries (W\,243,
W\,299, W\,400) and secondaries (W\,175, W\,311), we only give a range
of possible mass because of the ambiguity discussed in the text. The
last column gives the average binary mass ratios and is left empty for
the non-physical systems. Because the flux ratio of the W\,299 system
is so close to unity in all filters, we {\it assumed} the mass ratio
to be $q=1$.}
\label{tab:secmass}
\end{table}

From this study, we confirm as likely members most of the companions
discovered in this study and not already discarded as background
objects. On the other hand, we consider the companions to W\,290 and
W\,400 to be likely background stars, bringing the total number of
suspected background stars to seven. As a by-product, we find mass
ratios ranging from unity down to $q\lesssim0.1$, with primary masses
in the range 3--40\,$M_\odot$. Only two companions in our sample
appear to be high-mass stars (W\,205\,B and W\,254\,B, photometric
spectral types O9 and B3 respectively), and over half of the binaries
have $q\lesssim0.2$. This suggests that low mass ratios are indeed
extremely common in high-mass binaries. A consideration of $3\,\sigma$
systems yields a median mass ratio as high as $q\sim0.5$, because very
faint companions have a higher probability of being due to random
pairing due to the increasing surface density toward fainter
objects. On the other hand, a significant number of low-mass
companions may be unaccounted for when such a conservative criterion
is applied. There is no evidence for a change in mass ratio
distribution with primary mass in our sample, although small number
statistics prevent any detailed analysis.

We finally note that among the five stars of our Main Sample that are
supposed to have M-type companions according to \citet{dewinter97}
based on optical/near-infrared spectral energy distribution fitting,
two appear to be single in our survey (W\,262 and W\,273), one other
we identify as a chance projection binary (W\,339), and the remaining
two have companions that were detected here but with intermediate- to
early-type stars (W\,299 and W\,188). Only the one background
companion seems to be red enough to be responsible for the observed
near-infrared excess, suggesting that the other two systems may be
triple systems or that the significant near-infrared excess of their
primaries has some other cause than the presence of a close companion.


\subsection{Overall binary frequency in \ng}
\label{subsec:totbf}

Among the 60 targets of our Main Sample, 18 are found to have a visual
companion within 1\farcs5, the systems being evenly spread throughout
the cluster. Of these, seven are rejected as projected background
objects by virtue of their infrared photometric properties and/or
apparent mass ratios. We are thus left with eleven 2\,$\sigma$
candidate binaries, which implies an observed binary frequency of
$18\pm6\,\%$ over the 200--3000\,AU separation range.  Restricting our
sample to stars with OB spectral type or to those located in the core
of the cluster does not significantly alter this estimate: the binary
frequencies in those two subsamples are $20\pm6\,\%$ and
$19\pm6\,\%$. Finally, a conservative lower limit to the binary
frequency among high-mass stars in {\ng} can be obtained by retaining
only the systems that are above the 3\,$\sigma$ confidence level, as
explained in Sect.\,\ref{subsubsec:prob}; for our Main Sample, that
lower limit is $10\pm4\,\%$.

An important limitation of our survey is closely related to the
separation-dependent detection limit: we do not detect as many very
close binaries as we do for wider systems. Note, however, that
the apparent dearth of binaries seen in Figure\,\ref{fig:detect} at
short separation could partly be the result of a geometrical effect:
the area of the annulus extending from 0\farcs1 to 0\farcs5 represents
only about 10\,\% of the total surveyed area. It would be valuable to
estimate a completeness correction similar to those derived in
low-mass binary surveys \citep{bouvier97,duchene99ic}. However, such
corrections require that we make various assumptions concerning the
binary mass ratio and orbital period distributions. While they are
fairly well known for solar-type binaries, no similar unbiased,
well-defined distributions have been obtained for OB stars thus
far. It is thus impractical for us to perform such a correction for
our survey of NGC 6611.

Restricting the separation range considered here to 0\farcs5--1\farcs5
(1000--3000\,AU), on the other hand, proves to be a useful step. In
this range, any binary with a flux ratio $\Delta K\lesssim3.5$ can be
detected, and the average detection limit over this separation range
is $\Delta K\approx5.5$. For a typical star in our sample (median
spectral type B1, $M=10\,M_\odot$), such flux ratios correspond to
secondary masses of about $1\,M_\odot$ and $0.1\,M_\odot$,
respectively. We were thus able to detect any binary with mass ratio
$q\geq0.1$, as well as a large fraction of lower-mass stellar
companions. Once restricted to OB stars to ensure sample homogeneity,
we find $BF_{OB}^{vis}=14\pm5\,\%$ and $7\pm3\,\%$ when counting
systems above the 2\,$\sigma$ and 3\,$\sigma$ confidence levels,
respectively.

It is not possible to evaluate properly the {\it total} binary
frequency of high-mass stars, due to the limited sensitivity of
current surveys and because binaries with separations larger than
$\sim10$\,AU but smaller than 200\,AU cannot be detected either
spectroscopically or by current imaging surveys. However, we can
obtain a rough estimate by complementing our visual binary frequency
with that of \citet{bosch99}, i.e., assuming that the proportion of
spectroscopic binaries that they found in {\ng} can be extended to our
whole sample. They found a minimum of three binaries out of ten stars;
a somewhat higher frequency is quoted by
\citet{mermilliod_garcia01}. These numbers imply a binary frequency
for high-mass stars in {\ng} of $BF_{OB}^{tot}\gtrsim45\,\%$. Even
though this estimate is affected by significant uncertainties due to
small number statistics, the true frequency is likely much higher,
given the narrow range considered in our survey, the limited
sensitivity of both surveys, and the existence of two additional
binary candidates suggested by \citet{bosch99}. It is thus very likely
that most, if not all, OB stars in {\ng} have at least one stellar
companion. Finally, W\,175 is identified as a hierarchical triple
system, comprised of two O stars forming a spectroscopic binary
\citep{bosch99} and accompanied by a distant, third, A--F component
that has now been discovered in this survey. 

In line with previous studies, these results point toward a larger
binary frequency among OB stars than in any solar-type binary
population. For instance, \citet{duq_may91} found a total binary
frequency in field G dwarfs of about $BF_G^{tot}\approx60\,\%$,
considering 10 orders of magnitude in orbital periods and all
companions down to $q=0.1$. This is only slightly higher than the
estimate presented above, although the latter is restricted in its
range of orbital periods, as explained above. Excluding those systems
in our survey having extremely low mass ratios, $q<0.1$, which may
bias the comparison, from our 2\,$\sigma$ systems, we find
$BF_{OB}^{vis}(q\geq0.1)=10\pm4\,\%$ while the corresponding value for
field G-stars is $BF_G^{vis}(q\geq0.1)\approx4\,\%$ in the
1000--3000\,AU range. This represents a marginally significant
overabundance of wide visual binaries among OB stars of a factor of
about 2.5 with respect to field G dwarfs. It is unlikely that the
actual excess is much larger since our survey should be complete in
these separation and mass ratio ranges, as discussed above.

The binary frequency we find in {\ng} can also be compared to previous
imaging surveys of other high-mass star populations, although the
later suffer from various selection biases and varying detection
limits. Uncomplete surveys of field OB stars by \citet{lindroos85},
\citet{abt90}, and \citet{mason98} consistently yielded binary
frequencies on the order of 7--10\,\% in the separation range
1000-3000\,AU. More recently, \citet{bouvier_corporon01} and
\citet{hubrig01} used adaptive optics imaging to search for close
companions among the Herbig AeBe and X-ray selected late-B stars,
respectively. The proportion of companions among these populations are
both about 15\,\% in the same separation range when considering only
secondaries that could be found with our detection limit, a value
similar to our finding in {\ng} although one might expect the X-ray
selected sample to be biased towards higher binary rates. The only
other population of clustered high-mass stars surveyed so far is the
Trapezium OB stars \citep{preibisch99}. Because it is much closer to
the Sun than {\ng}, the separation range common to both surveys is
restrained to 200--800\,AU over which our {\ng} images have a very
limited detection limit. From their sample of 13 targets, only two
companions dectected by \citeauthor{preibisch99} would have been
detected if the Trapezium was at the same distance as {\ng}, yielding
a restricted binary frequency that is not significantly different from
our findings over that separation range.



To summarize, we conclude that the frequency of wide companions among
high-mass stars in {\ng} is similar to that observed in several other
populations. A high proportion of spectroscopic binaries is observed
in both {\ng} and the field populations \citep{abt90}: the 30\,\%
frequency of spectroscopic binaries in {\ng} revealed by
\citet{bosch99}, once again limited by the small size of their sample,
is already at least comparable to that of the field G dwarfs, which
host a proportion of binaries that is about 22\,\% for periods shorter
than 30\,years \citep{duq_may91}. Overall, there is a convergent array
of evidence to suggest that the binary rate is enhanced in OB stars by
comparison with solar-type stars in the field, independent of the
environment in which these massive stars are found (whether clusters
or as isolated field stars) although none of the individual studies
allows a firm conclusion due to small sample sizes, inhomogeneous
samples and/or selection biases.


\subsection{Other binary properties}
\label{subsec:binprop}

Correlations between binary properties and particular stellar
characteristics, especially stellar mass, may be of equal or even
potentially greater significance than the raw binarity rate itself.
\citet{preibisch99} suggested that the most massive stars in the
Trapezium cluster host more companions than do the stars later than
spectral type B3 or so. However, this trend vanishes when one
considers only the companions that are at least 200\,AU from their
primaries, i.e., at separations similar to those probed by our survey,
which shows no evidence of such a trend. Indeed, we can split our
sample into two roughly equal subsamples: 29 stars of spectral type B1
or earlier, and 25 stars of later spectral type. We then find no
statistical difference between the binary frequencies of those
subsamples ($21\pm8$\,\% and $16\pm8$\,\%).

Most of the binaries detected in our survey have small mass ratios, 
in good agreement with results for field star surveys
\citep{abt90,mason98}. The latter studies, as well as the
\citet{preibisch99} study of the Trapezium, concluded that the
mass ratios of visual binaries can be reproduced by assuming the
secondary masses are taken at random from the universal initial mass
function (IMF). Given our inefficiency in detecting very low-mass
ratio systems and our statistical limitations, we cannot confirm
whether this conclusion applies to {\ng}. On the other hand, the mass
ratio distribution for high-mass {\it spectroscopic} binaries appears
to be much flatter \citep[e.g.,][]{mason98}, which suggests that two
different formation processes are involved; however, spectroscopic
surveys are known to be relatively inefficient in detecting low-mass
secondaries. In conclusion, the IMF distribution of secondary masses
seems reasonable for visual binaries in all samples studied so far. We
also note that, although long-period solar-type binaries show an
almost flat mass ratio distribution, the latter is consistent with an
IMF pairing of companions \citep{duq_may91} (at least down to
$q=0.2$), which in turn may suggest that the formation mechanism of
binary systems is not dramatically different for low- and high-mass
stars.

Among other properties that could be linked to binarity is the line
emission in the spectra of these massive stars. Emission lines have
been identified in many of the B stars of this cluster, and we
identify 28 emission stars in our sample from the results of
\citet{the90}, \citet{hillenbrand93} and \citet{dewinter97}
\citep[for a contrary view, c.f.][]{herbig_dahm01}. Four of these stars
were found to have companions in our survey, representing a binary
frequency of $14\pm7\,\%$ among this subsample, which is
indistinguishable from the frequency derived for the sample as a
whole. It thus seems that the presence of a wide companion is
unrelated to the emission properties of B type stars in {\ng}.

Finally, through a full coverage of the optical and near-infrared
domains, \citet{hillenbrand93} and \citet{dewinter97} have identified
stars possessing significant near-infrared excess. Such excesses can
be related to the presence of a circumstellar disk, to an abnormal
extinction law, or to the existence of a close red M-type companion to
the stars \citep{dewinter97}. Although the exact origin of this excess
is unknown in many case, it is interesting to note that the proportion
of binary systems among stars having a near-infrared excess is similar
to that of the full sample (two binaries among twelve targets, or
$\sim17\,\%$). The presence of a companion thus seems independent of
any source of near-infrared excess, such as a circumstellar disk.


\section{Implication for the formation mechanism of high-mass stars}
\label{sec:discus}

The main finding of our survey is that wide binaries are rather common
among massive stars in a cluster like {\ng}. In the framework of the
model presented by \citet{bonnell98}, which seeks to account for the
formation of massive stars through numerous mergers, this result
raises at least two related questions: How exactly do these systems
form? How do they survive for at least a couple of million years
despite the many close-by interactions they have with other cluster
members?  By comparison, the accretion process suggested by
\citet{nakano89}, which takes place through an equatorial disk
and assumes that each massive protostar is relatively isolated, has
little difficulty retaining companions once they are formed. How such
configurations would form to start with nonetheless remains
unclear. In this section, we first consider the issue of the survival
of wide binaries in dense stellar environments, and then we try to
understand how such systems fit into the framework of our current
knowledge of massive star formation.


\subsection{Survival of wide massive binaries in dense clusters}
\label{subsec:surviv}

The disruptive effects of the dynamical evolution of dense clusters
have been studied in detail through numerical simulations over the
last years \citep{kroupa95c,kroupa99,kroupa01}. Although these works
were primarily interested in the evolution of low-mass binaries, some
general conclusions can be extended to high-mass stars. First, the
timescale over which wide binaries are significantly depleted is the
cluster crossing time. Second, markedly different behaviours are
expected for ``soft'' and ``hard'' binaries, the division between the
two types being those systems whose binding energy is equal to the
average (translational) kinetic energy of the cluster constituents.

The crucial parameters here, the crossing time and the velocity
dispersion, are currently unknown in {\ng}, both because the low-mass
population, hence the total mass of the cluster, has not yet been
probed, and because of the very small proper motion of the cluster
members. Still, we can derive a rough estimate of the soft/hard
boundary if we arbitrarily assume that the velocity dispersion of
{\ng} is similar to that of the Trapezium cluster ($\sigma\sim2\,{\rm
km\,s}^{-1}$), which is the best known massive cluster of comparable
age.

\citet{kroupa95c} has shown that the distinction between soft and
hard systems in the Trapezium cluster occurs at separations of the
order 1000\,AU for solar-type primaries; a striking lack of systems
wider than this limit has been subsequently documented by
\citet{scally99}. Since in our survey we consider primaries that are
on average 10 times more massive, this limit is pushed outwards to
about 3500\,AU in {\ng} because of the deeper potential well. This
value is somewhat larger than the widest systems we measure
here. However, according to the model results, systems with 
separations that are a significant fraction of this limit should 
also be greatly depleted in a few crossing times, although not 
completely \citep{kroupa01}.

Qualitatively, we can say that binaries with separations as large as
3000\,AU can survive in an environment as dense as the Trapezium
cluster or {\ng}, although some of the primordial systems may already
have been destroyed.  Provided wide system can form during the
high-mass star formation process, it is thus not surprising that they
are able to live for a few million years. Even so, it is likely that
there were even more wide massive binaries early on in the cluster
history than exist now, and this may be harder yet to reconcile with
the merger model, as we discuss below.


\subsection{High-mass star formation: merging versus accretion
processes}

The model proposed by \citet{bonnell98} to explain the formation of
massive stars is based on dramatic dynamical events, which involve
mergers of intermediate mass protostellar cores. In this framework,
most of the mergers would occur before the cores have contracted into
stellar objects, at an early epoch in the cluster history when the
environment is extremely dense. Nevertheless, most two-body
interactions do not lead to mergers but rather to close
encounters. The latter can form binary systems through the influence
of tidal forces, which may be strong enough to brake a grazing
fragment down to a bound orbit if its periastron is only a few stellar
radii, as described by \citet{bonnell98}. Then, because these
protobinaries form very early in the process, subsequent accretion
onto the system will likely feed both seeds at similar rates and yield
a tight binary that is comprised of two relatively massive stars. The
terminal state at the end of this process qualitatively matches the
observations of high-mass spectroscopic binaries described above.
However, tidal interactions would not appear to be efficient enough to
form binaries as wide as the visual binaries in our survey, and so an
explanation for the existence of those binaries is still needed.

Consider the following picture. Let us assume that for a massive star
to form as the product of tidal interactions, the stellar density
during the merging process was necessarily much higher than it is now
(so as to ensure a large enough number of mergers). Therefore, any
wide binary formed early on in the cluster history would have been
rapidly disrupted, and the systems that are still present in the
cluster must have formed after the density dropped significantly,
i.e., after the merging period ended. One possibility is that a
fragment might be captured at the time of the last merger as a result
of a dissipative three-body interaction. In that case, the mass of the
wide companions would roughly follow that of the fragments, hence the
IMF, which is not inconsistent with the observations. Of course, this
is an {\it ad hoc} explanation and it may be unrealistic to assume
that encounters and mergers stopped so abruptly unless gas removal
from the cluster core occurs on a timescale at least as short as the
merger frequency.

Another possibility is that tidal interactions during mergers and/or
grazing encounters are able to disrupt the fragments before they
merge, and thus lead to the formation of a significant accretion disk
around the massive protostar. Subsequently this disk might
disintegrate into smaller fragments and in the process form a low-mass
companion. Observational evidence shows that disks as large as several
thousand AU do exist \citep[e.g.,][]{odell_wen94,vandertak00}. Due to
their spatial extent and viscous properties, disks are more difficult
to sweep out completely through a nearby encounter than it is to eject
a similarly massive companion. Indeed, the perturbation induced by an
object that is passing by could trigger the fragmentation of the disk
into a secondary component. Hence, the disks could survive for a long
time, even if they form in the early phases of the cluster
creation. However, the frequency of such large disks is currently
unknown and may be too low to account for the observed large
proportion of wide companions.

Overall, the violent process suggested by \citet{bonnell98} to
describe the formation of massive stars agrees with the observed
properties of spectroscopic binaries, but presents some difficulties
with regard to much wider systems. Another difficulty is the
similarity in the binary properties of the B stars in the cluster and
field populations. While the former would have experienced mergers,
the latter are more likely to have formed via accretion in quieter and
looser environments, since most of them are not identified as runaway
stars from some massive cluster and therefore probably formed from a
small molecular cloud. Such different formation paths would appear
unlikely to yield a similar binary frequency and properties, casting
some doubt upon the model described above.

To circumvent these difficulties, let us consider the possibility 
that high-mass objects form through the canonical accretion process
\citep{beech_mitalas94}, even in dense clusters. In this scenario,
high-mass stars are surrounded by accretion disks which, again, may
fragment into a stellar companion. The disks need to have a mass that
is comparable to that of the central star in order to account for both
the triggering of the gravitational instability and the existence of
massive companions in such an accretion scenario. Subsequent accretion
would again probably lead to the formation of a system consisting of
two high-mass stars as the disk would likely become unstable very
fast. However, the early fragmentation of the disk would prevent
the central object from reaching a high mass by halting the viscous
phenomenon. Similarly, binaries could form through the fission of a
collapsing envelope, before radiation pressure dissipates it. Again,
this would occur very early in the formation process and would lead to
equal mass systems for the same reasons as above. The large number of
unequal mass systems does not seem to fit very well with both of these
models.

Alternatively, visual binaries could simply result from the
fragmentation of the molecular cloud into two bound, widely separated
prestellar cores that would subsequently collapse, as is supposed to
be the case for low-mass binaries. Indeed, the objects have to be
fairly isolated for the accretion process to proceed until the star
reaches a mass $M>10M_\odot$. Therefore, this model requires that the
stellar density has not been much higher in the past than it is now,
in which case one does not expect strong dynamical effects from
mergers or very close encounters that could disrupt these wide
systems. In this picture, cloud fragmentation naturally leads to a
companion mass distribution that is similar to the IMF, as
observations suggest for both high- and low-mass field stars. As a
consequence of this universal scenario, one does not expect to find
dramatically different binary properties between O and B stars, again
an expectation that is in agreement with the observations. The
overabundance of wide binaries among high-mass stars would simply be
the tracer of their enhanced capability to retain low-mass companions
through their deeper potential well (see
Sect.\,\ref{subsec:surviv}). This would explain why both cluster and
field massive binaries present similar properties while the binary
frequency of solar-type populations appears to be dependent upon
environmental conditions and particularly the stellar density
\citep{duchene99ic,bouvier01}.

Although the formation process through residual gas accretion seems to
match more naturally the observed massive binary properties, it is not
without its own difficulties. Most importantly, one might wonder
whether high-mass stars that form in a rich cluster such as {\ng} can
be considered to be isolated throughout the accretion process, which
lasts for a million years or so
\citep{beech_mitalas94,behrend_maeder01}. This may be correct if {\ng}
has had constant properties since its birth, and if these
characteristics are similar to those of the Orion Trapezium cluster in
its current stage. However, studies of the dynamical evolution of
stellar clusters show that gas removal in the first few millions years
causes the cluster to expand significantly
\citep[e.g.][]{kroupa99}. If {\ng} has actually been an order of
magnitude or more denser than it currently is, it would seem hard to
disregard the merging process as an important channel for high-mass
star formation and to keep considering protostars as isolated from the
rest of the cluster in which they are embedded.


\section{Conclusion}
\label{sec:concl}

We have used the adaptive optics system at CFHT to survey 60 high-mass
stars, most of them of spectral type O or B, to look for companions in
the projected separation range 200--3000\,AU
(0\farcs1--1\farcs5). This is the first visual binary survey among
such a large and homogeneous sample of high-mass stars. We identified
11 likely physical systems with flux ratios as large as $\Delta
K\gtrsim5\,$mag. This yields an uncorrected binary frequency of
$18\pm6\,\%$ in the orbital period range $10^3\lesssim P({\rm
yrs})\lesssim5\times10^4$. Limiting our search to binaries with mass
ratios $q\geq0.1$, we find a marginally significant overabundance of
companions among OB stars in {\ng} than in the field population of
solar-type dwarfs, reinforcing the trend that high-mass stars host
more wide companions than do lower-mass objects, both in the visual
and spectroscopic domains.

Mass ratios have been derived for every system whose primary spectral 
type is known.  It appears that half of the detected binaries have
$q\lesssim0.2$, pointing to a large predominance of small mass ratios
in wide massive binaries. This behaviour is in rough agreement with
previous surveys, which have generally favored a random association of
companions from the initial mass function with massive primaries. No
obvious correlation is found between binarity and primary mass,
propensity toward emission lines, or infrared excess of the system. 

The high rate of occurence of wide systems (several hundred AU) among
massive stars seems to contradict the formation model based on mergers
that has been proposed by \citet{bonnell98}. Although this model
qualitatively predicts the properties and high frequency of {\it
spectroscopic} binaries, the numerous encounters between protostars in
the early stages of the cluster evolution should succeed in disrupting
most of the wide visual binaries, due to their lower binding energy. To
reconcile this model with the observed visual binary frequency, the
capture of a low-mass wide companion has to occur at the time of the
last merging event, so that the number of subsequent interactions with
other objects is small. Another possible interpretation is that an
extended disk of material is created and fed by successive mergers and
tight encounters, which would subsequently fragment to form a wide
secondary. Both of these explanations are not very satisfactory, as
they require that a large number of close encounters transpire within
a very narrow range of physical parameters such as impact parameter,
core sizes, masses, etc.

An alternative proposed model to account for the formation of massive
stars, based on large disk accretion on the protostar, arouses fewer
objections, as wide binaries would likely result from the
fragmentation of the cloud itself. However, for the accretion process
to last long enough and to result in a high-mass star, each system
must be substantially isolated for at least a million years or
so. Although this might well be the correct picture for field OB
stars, it is likely that clusters are very much denser in their
initial formation stages, when close encounters and maybe even mergers
can play an important role in the evolution of the protostars within
such regions. One way to reconcile dense environments and disk
accretion processes would be, for future numerical simulations, to
show that close encounters, although they strongly perturb
circumstellar disks, do not shorten significantly their lifetime. 

From an observational approach, the detection of numerous large
circumstellar disks (with radii of several hundreds of AU) around OB
stars is an expected outcome of future deep imaging programs if these
objects do form through continous accretion processes. On the other
hand, high-angular resolution observations of deeply embedded
intermediate- and high-mass protostars are awaited to test the mergers
model, in which high-mass stars should be deficient in very young
clusters, before most of the mergers have occurred.

\begin{acknowledgements}
We thank J.-L. Beuzit for obtaining a subset of the data presented in
this paper and L. Siess for valuable exchanges regarding evolutionary
tracks. We also thank A. Ghez and C. McCabe for obtaining a Keck
spectrum of one of the suspected background companions, as well as for
critical comments about the paper. A prompt report from an anonymous
referee helped to improve and clarify this manuscript. J.E. and
J.B. acknowledge financial support through a PROCOPE exchange
programme by APAPE and DAAD. G.D. was partially supported at UCLA by
the NSF Science and Technology Center for Adaptive Optics, managed by the 
University of California at Santa Cruz under cooperative agreement No.
AST-9876783.
\end{acknowledgements}

\small

\bibliographystyle{apj}
\bibliography{biblio.bib}

\begin{thebibliography}{55}
\expandafter\ifx\csname natexlab\endcsname\relax\def\natexlab#1{#1}\fi

\bibitem[{{Abt}(1988)}]{abt88}
{Abt}, H.~A. 1988, ApJ, 331, 922

\bibitem[{{Abt} {et~al.}(1990){Abt}, {Gomez}, \& {Levy}}]{abt90}
{Abt}, H.~A., {Gomez}, A.~E., \& {Levy}, S.~G. 1990, ApJS, 74, 551

\bibitem[{{Beech} \& {Mitalas}(1994)}]{beech_mitalas94}
{Beech}, M. \& {Mitalas}, R. 1994, ApJS, 95, 517

\bibitem[{{Behrend} \& {Maeder}(2001)}]{behrend_maeder01}
{Behrend}, R. \& {Maeder}, A. 2001, A\&A, 373, 190

\bibitem[{{Belikov} {et~al.}(1999){Belikov}, {Kharchenko}, {Piskunov}, \&
  {Schilbach}}]{belikov99}
{Belikov}, A.~N., {Kharchenko}, N.~V., {Piskunov}, A.~E., \& {Schilbach}, E.
  1999, A\&AS, 134, 525

\bibitem[{{Belikov} {et~al.}(2000){Belikov}, {Kharchenko}, {Piskunov}, \&
  {Schilbach}}]{belikov00}
---. 2000, A\&A, 358, 886

\bibitem[{{B\"ohm-Vitense}(1981)}]{bohm81}
{B\"ohm-Vitense}, E. 1981, ARA\&A, 19, 295

\bibitem[{{Bonnell} {et~al.}(1998){Bonnell}, {Bate}, \&
  {Zinnecker}}]{bonnell98}
{Bonnell}, I.~A., {Bate}, M.~R., \& {Zinnecker}, H. 1998, MNRAS, 298, 93

\bibitem[{{Bonnell} \& {Davies}(1998)}]{bonnell_davies98}
{Bonnell}, I.~A. \& {Davies}, M.~B. 1998, MNRAS, 295, 691

\bibitem[{{Bosch} {et~al.}(1999){Bosch}, {Morrell}, \& {Niemela}}]{bosch99}
{Bosch}, G.~I., {Morrell}, N.~I., \& {Niemela}, V.~S. 1999, RMxAA, 35, 85

\bibitem[{{Bouvier} \& {Corporon}(2001)}]{bouvier_corporon01}
{Bouvier}, J. \& {Corporon}, P. 2001, in {\it The formation of binary stars},
  ed. {Mathieu} \& {Zinnecker}, Vol. 200, ASP conf. series, 155

\bibitem[{{Bouvier} \& {Duch\^ene}(2001)}]{bouvier_duchene01}
{Bouvier}, J. \& {Duch\^ene}, G. 2001, in {\it The Origins of Stars and
  Planets: The VLT View}, ed. {Alves} \& {McCaughrean}, ESO Astrophysics
  Symposia

\bibitem[{{Bouvier} {et~al.}(2001){Bouvier}, {Duch\^ene}, {Mermilliod}, \&
  {Simon}}]{bouvier01}
{Bouvier}, J., {Duch\^ene}, G., {Mermilliod}, J.-C., \& {Simon}, T. 2001, A\&A,
  375, 989

\bibitem[{{Bouvier} {et~al.}(1997){Bouvier}, {Rigaut}, \& {Nadeau}}]{bouvier97}
{Bouvier}, J., {Rigaut}, F., \& {Nadeau}, D. 1997, A\&A, 323, 139

\bibitem[{{de Winter} {et~al.}(1997){de Winter}, {Kouis}, {Th\'e}, {van Den
  Ancker}, {P\'erez}, \& {Bibo}}]{dewinter97}
{de Winter}, D., {Kouis}, C., {Th\'e}, P.~S., {van Den Ancker}, M.~E.,
  {P\'erez}, M.~R., \& {Bibo}, E.~A. 1997, A\&AS, 121, 223

\bibitem[{{Doyon} {et~al.}(1998){Doyon}, {Nadeau}, {Vall\'ee}, {Starr},
  {Cuillandre}, {Beuzit}, {Beigbeder}, \& {Brau-Nogue}}]{doyon98}
{Doyon}, R., {Nadeau}, D., {Vall\'ee}, P., {Starr}, B.~M., {Cuillandre}, J.-C.,
  {Beuzit}, J.-L., {Beigbeder}, F., \& {Brau-Nogue}, S. 1998, Proc. SPIE, 3354,
  760

\bibitem[{{Duch\^ene} {et~al.}(1999){Duch\^ene}, {Bouvier}, \&
  {Simon}}]{duchene99ic}
{Duch\^ene}, G., {Bouvier}, J., \& {Simon}, T. 1999, A\&A, 343, 831

\bibitem[{{Duncan}(1920)}]{duncan20}
{Duncan}, J.~C. 1920, ApJ, 51, 4

\bibitem[{{Duquennoy} \& {Mayor}(1991)}]{duq_may91}
{Duquennoy}, A. \& {Mayor}, M. 1991, A\&A, 248, 485

\bibitem[{{Garmany} {et~al.}(1980){Garmany}, {Conti}, \& {Massey}}]{garmany80}
{Garmany}, C.~D., {Conti}, P.~S., \& {Massey}, P. 1980, ApJ, 242, 1063

\bibitem[{{Ghez} {et~al.}(1993){Ghez}, {Neugebauer}, \& {Matthews}}]{ghez93}
{Ghez}, A.~M., {Neugebauer}, G., \& {Matthews}, K. 1993, AJ, 106, 2005

\bibitem[{{Gieseking}(1982)}]{gieseking82}
{Gieseking}, F. 1982, A\&AS, 49, 673

\bibitem[{{Herbig} \& {Dahm}(2001)}]{herbig_dahm01}
{Herbig}, G.~H. \& {Dahm}, S.~E. 2001, PASP, 113, 195

\bibitem[{{Hillenbrand} \& {Hartmann}(1998)}]{hillen_hartmann98}
{Hillenbrand}, L.~A. \& {Hartmann}, L.~W. 1998, ApJ, 492, 540

\bibitem[{{Hillenbrand} {et~al.}(1993){Hillenbrand}, {Massey}, {Strom}, \&
  {Merrill}}]{hillenbrand93}
{Hillenbrand}, L.~A., {Massey}, P., {Strom}, S.~E., \& {Merrill}, K.~M. 1993,
  AJ, 106, 1906

\bibitem[{{Hubrig} {et~al.}(2001){Hubrig}, {Le Mignant}, {North}, \&
  {Krautter}}]{hubrig01}
{Hubrig}, A., {Le Mignant}, D., {North}, P., \& {Krautter}, J. 2001, A\&A, 372,
  152

\bibitem[{{Kharchenko} \& {Schilbach}(1995)}]{kharchenko_schilbach95}
{Kharchenko}, N. \& {Schilbach}, E. 1995, Astronomische Nachrichten, 316, 91

\bibitem[{{Kroupa}(1995)}]{kroupa95c}
{Kroupa}, P. 1995, MNRAS, 277, 1522

\bibitem[{{Kroupa} {et~al.}(2001){Kroupa}, {Aarseth}, \& {Hurley}}]{kroupa01}
{Kroupa}, P., {Aarseth}, S., \& {Hurley}, J. 2001, MNRAS, 321, 699

\bibitem[{{Kroupa} {et~al.}(1999){Kroupa}, {Petr}, \& {McCaughrean}}]{kroupa99}
{Kroupa}, P., {Petr}, M.~G., \& {McCaughrean}, M.~J. 1999, New A, 4, 495

\bibitem[{{Leinert} {et~al.}(1993){Leinert}, {Zinnecker}, {Weitzel},
  {Christou}, {Ridgway}, {Jameson}, {Haas}, \& {Lenzen}}]{leinert93}
{Leinert}, C., {Zinnecker}, H., {Weitzel}, N., {Christou}, J., {Ridgway},
  S.~T., {Jameson}, R., {Haas}, M., \& {Lenzen}, R. 1993, A\&A, 278, 129

\bibitem[{{Lindroos}(1985)}]{lindroos85}
{Lindroos}, K.~P. 1985, A\&AS, 60, 183

\bibitem[{{Mason} {et~al.}(1998){Mason}, {Gies}, {Hartkopf}, {Bagnuolo},
  {Brummelaar}, \& {McAlister}}]{mason98}
{Mason}, B.~D., {Gies}, D.~R., {Hartkopf}, W.~I., {Bagnuolo}, W.~G.,
  {Brummelaar}, T.~T., \& {McAlister}, H.~A. 1998, AJ, 115, 821

\bibitem[{{McAlister} {et~al.}(1993){McAlister}, {Mason}, {Hartkopf}, \&
  {Shara}}]{mcalister93}
{McAlister}, H.~A., {Mason}, B.~D., {Hartkopf}, W.~I., \& {Shara}, M.~M. 1993,
  AJ, 106, 1639

\bibitem[{{McBreen} {et~al.}(1982){McBreen}, {Fazio}, \& {Jaffe}}]{mcbreen82}
{McBreen}, B., {Fazio}, G.~G., \& {Jaffe}, D.~T. 1982, ApJ, 254, 126

\bibitem[{{Meaburn} \& {White}(1982)}]{meaburn_white82}
{Meaburn}, J. \& {White}, N.~J. 1982, MNRAS, 199, 121

\bibitem[{{Mermilliod} \& {Garc\'{\i}a}(2001)}]{mermilliod_garcia01}
{Mermilliod}, J.-C. \& {Garc\'{\i}a}, B. 2001, in {\it The formation of binary
  stars}, ed. {Mathieu} \& {Zinnecker}, Vol. 200, ASP conf. series, 191

\bibitem[{{Morrell} \& {Levato}(1991)}]{morrell_levato91}
{Morrell}, N. \& {Levato}, H. 1991, ApJS, 75, 965

\bibitem[{{Nakano}(1989)}]{nakano89}
{Nakano}, T. 1989, ApJ, 345, 464

\bibitem[{{O'Dell} \& {Wen}(1994)}]{odell_wen94}
{O'Dell}, C.~R. \& {Wen}, Z. 1994, ApJ, 436, 194

\bibitem[{{Petr} {et~al.}(1998){Petr}, {Coud\'e Du Foresto}, {Beckwith},
  {Richichi}, \& {McCaughrean}}]{petr98}
{Petr}, M.~G., {Coud\'e Du Foresto}, V., {Beckwith}, S. V.~W., {Richichi}, A.,
  \& {McCaughrean}, M.~J. 1998, ApJ, 500, 825

\bibitem[{{Preibisch} {et~al.}(1999){Preibisch}, {Balega}, {Hofmann},
  {Weigelt}, \& {Zinnecker}}]{preibisch99}
{Preibisch}, T., {Balega}, Y., {Hofmann}, K., {Weigelt}, G., \& {Zinnecker}, H.
  1999, New A, 4, 531

\bibitem[{{Prosser} {et~al.}(1994){Prosser}, {Stauffer}, {Hartmann},
  {Soderblom}, {Jones}, {Werner}, \& {McCaughrean}}]{prosser94}
{Prosser}, C.~F., {Stauffer}, J.~R., {Hartmann}, L., {Soderblom}, D.~R.,
  {Jones}, B.~F., {Werner}, M.~W., \& {McCaughrean}, M.~J. 1994, ApJ, 421, 517

\bibitem[{{Rieke} \& {Lebofsky}(1985)}]{rieke_lebof85}
{Rieke}, G.~H. \& {Lebofsky}, M.~J. 1985, ApJ, 288, 618

\bibitem[{{Rigaut} {et~al.}(1998){Rigaut}, {Salmon}, {Arsenault}, {Thomas},
  {Lai}, {Rouan}, {V\'eran}, {Gigan}, {Crampton}, {Fletcher}, {Stilburn},
  {Boyer}, \& {Jagourel}}]{rigaut98}
{Rigaut}, F., {Salmon}, D., {Arsenault}, R., {Thomas}, J., {Lai}, O., {Rouan},
  D., {V\'eran}, J.-P., {Gigan}, P., {Crampton}, D., {Fletcher}, J.~M.,
  {Stilburn}, J., {Boyer}, C., \& {Jagourel}, P. 1998, PASP, 110, 152

\bibitem[{{Scally} {et~al.}(1999){Scally}, {Clarke}, \&
  {McCaughrean}}]{scally99}
{Scally}, A., {Clarke}, C., \& {McCaughrean}, M.~J. 1999, MNRAS, 306, 253

\bibitem[{{Schaller} {et~al.}(1992){Schaller}, {Schaerer}, {Meynet}, \&
  {Maeder}}]{schaller92}
{Schaller}, G., {Schaerer}, D., {Meynet}, G., \& {Maeder}, A. 1992, A\&AS, 96,
  269

\bibitem[{{Shu} {et~al.}(1987){Shu}, {Adams}, \& {Lizano}}]{shu87}
{Shu}, F.~H., {Adams}, F.~C., \& {Lizano}, S. 1987, ARA\&A, 25, 23

\bibitem[{{Siess} {et~al.}(2000){Siess}, {Dufour}, \& {Forestini}}]{siess00}
{Siess}, L., {Dufour}, E., \& {Forestini}, M. 2000, A\&A, 358

\bibitem[{{Simon} {et~al.}(1995){Simon}, {Ghez}, {Leinert}, {Cassar}, {Chen},
  {Howell}, {Jameson}, {Matthews}, {Neugebauer}, \& {Richichi}}]{simon95}
{Simon}, M., {Ghez}, A.~M., {Leinert}, C., {Cassar}, L., {Chen}, W.~P.,
  {Howell}, R.~R., {Jameson}, R.~F., {Matthews}, K., {Neugebauer}, G., \&
  {Richichi}, A. 1995, ApJ, 443, 625

\bibitem[{{Th\'e} {et~al.}(1990){Th\'e}, {de Winter}, {Feinstein}, \&
  {Westerlund}}]{the90}
{Th\'e}, P.~S., {de Winter}, D., {Feinstein}, A., \& {Westerlund}, B.~E. 1990,
  A\&AS, 82, 319

\bibitem[{{van der Tak} {et~al.}(2000){van der Tak}, {van Dishoeck}, {Evans},
  \& {Blake}}]{vandertak00}
{van der Tak}, F. F.~S., {van Dishoeck}, E.~F., {Evans}, N.~J., \& {Blake},
  G.~A. 2000, ApJ, 537, 283

\bibitem[{{van Dessel} \& {Sinachopoulos}(1993)}]{vandessel_sinachopoulos93}
{van Dessel}, E. \& {Sinachopoulos}, D. 1993, A\&AS, 100, 517

\bibitem[{{Walker}(1961)}]{walker61}
{Walker}, M.~F. 1961, ApJ, 133, 438

\bibitem[{{Wolfire} \& {Cassinelli}(1987)}]{wolfire_cassinelli87}
{Wolfire}, M.~G. \& {Cassinelli}, J.~P. 1987, ApJ, 319, 850

\end{thebibliography}

\end{document}